\documentclass{desyproc}
\usepackage{url,cite,wrapfig}

\begin{document}
%------------------------------------
\title{Hunting magnetic monopoles and more with \\ MoEDAL at the LHC}

%for single authors the superscripts are optional
\author{{\slshape Vasiliki A.\ Mitsou$^1$  on behalf of the MoEDAL Collaboration}\\[1ex]
$^1$Instituto de F\'isica Corpuscular (IFIC), CSIC -- Universitat de Val\`encia,
Valencia, Spain \\ }

% if the proceedings are available online (e.g. at Indico)
% please enter the contribution ID or file_name below for the DOI
%\contribID{32}
\contribID{83}

% TO THE CONFERENCE EDITORS: 
% please update the following information      
% before sending the template to the authors
\confID{13889}  % if the conference is on Indico uncomment this line
\desyproc{DESY-PROC-2017-XX}
\acronym{Patras 2017} % if you want the Acronym in the page footer uncomment this line
\doi  % if there is an online version we will register DOIs

\maketitle

\begin{abstract}
The MoEDAL experiment at the LHC is optimised to detect highly-ionising particles such as magnetic monopoles, dyons and (multiply) electrically-charged stable massive particles predicted in a number of theoretical scenarios. MoEDAL, deployed in the LHCb cavern, combines passive nuclear track detectors with magnetic monopole trapping volumes, while backgrounds are being monitored with an array of MediPix detectors. The detector concept and its physics reach is presented with emphasis given to recent results on monopoles.  \end{abstract}

%%%%%%%%%%%%%%%%%%%%%%%%%%%%%%%%%%%%%%%%%%%%%%%%%%%
%%%%%%%%%%%%%%%%%%%%%%%%%%%%%%%%%%%%%%%%%%%%%%%%%%%
\section{Introduction}\label{sc:intro}

MoEDAL (Monopole and Exotics Detector at the LHC)~\cite{moedal}, the $7^{\rm th}$ experiment to operate at the Large Hadron Collider (LHC), is designed to search for manifestations of new physics through highly-ionising (HI) particles in a manner complementary to ATLAS and CMS~\cite{DeRoeck:2011aa}. The main motivation for the MoEDAL experiment is to pursue the quest for magnetic monopoles at LHC energies. Nonetheless the detector is also designed to search for any massive,  long-lived, slow-moving particle~\cite{Fairbairn07} with single or multiple electric charges arising in many scenarios of physics beyond the Standard Model~\cite{Acharya:2014nyr}. 

%%%%%%%%%%%%%%%%%%%%%%%%%%%%%%%%%%%%%%%%%%%%%%%%%%%
%%%%%%%%%%%%%%%%%%%%%%%%%%%%%%%%%%%%%%%%%%%%%%%%%%%
\section{The MoEDAL detector}\label{sc:detector}

\begin{wrapfigure}[15]{O}{0.52\textwidth}
  \centering
  \includegraphics[width=0.5\textwidth]{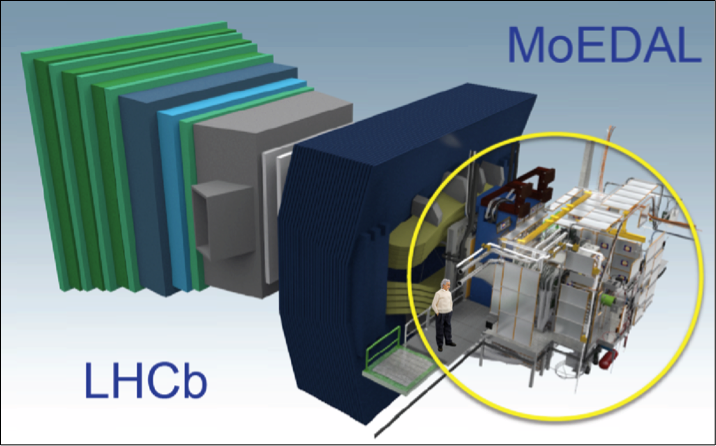}
  \caption{\label{fg:moedal-lhcb} A three-dimensional schematic view of the MoEDAL detector (in the yellow circle) around the LHCb VELO region at Point~8 of the LHC.}
\end{wrapfigure}

The MoEDAL detector~\cite{moedal} is deployed around the intersection region at the LHC Point~8 in the LHCb Vertex Locator (VELO) cavern. A schematic view of the MoEDAL experiment is shown in Fig.~\ref{fg:moedal-lhcb}. It is a unique and largely passive detector comprising different detector technologies. 

%%%%%%%%%%%%%%%%%%%%%%%%%%%%%%%%%%%%%%%%%%%%%%%%%%%
%\subsection{Low-threshold nuclear track detectors}\label{sc:ndt}
\subsection{Nuclear track detectors}\label{sc:ndt}

The main sub-detector system is made of a large array of CR39$^{\textregistered}$,  Makrofol$^{\textregistered}$ and Lexan$^{\textregistered}$ nuclear track detector (NTD) stacks surrounding the intersection area. The passage of a HI particle through the plastic detector is marked by an invisible damage zone along the trajectory. The damage zone is revealed as a cone-shaped etch-pit when the plastic detector is chemically etched. Then the sheets of plastics are scanned looking for aligned etch pits in multiple sheets. The MoEDAL NTDs have a threshold of $Z/\beta\sim5$, where $Z$ is the charge and $\beta=v/c$ the velocity of the incident particle. 

%%%%%%%%%%%%%%%%%%%%%%%%%%%%%%%%%%%%%%%%%%%%%%%%%%%
%\subsection{Very high-charge catcher}\label{sc:vhcc}

Another type of NTD installed is the Very High Charge Catcher ($Z/\beta\sim50$). It consists of two flexible low-mass stacks of Makrofol$^{\textregistered}$, deployed in the LHCb acceptance between RICH1 and the Trigger Tracker. It is the only NTD (partly) covering the forward region, adding only $\sim0.5\%$ to the LHCb material budget while enhancing considerably the overall geometrical coverage of MoEDAL.

%%%%%%%%%%%%%%%%%%%%%%%%%%%%%%%%%%%%%%%%%%%%%%%%%%%
\subsection{Magnetic trappers}\label{sc:mmt}

A unique feature of the MoEDAL detector is the use of paramagnetic magnetic monopole trappers (MMTs) to capture magnetically-charged HI particles. The aluminium absorbers of MMTs are subject to an analysis looking for magnetically-charged particles at a remote SQUID magnetometer facility~\cite{Joergensen:2012gy}. For the 2015 run at 13~TeV, the MMT consisted of 672~aluminium rods for a total mass of 222~kg that were placed 1.62~m from the IP8 LHC interaction point under the beam pipe on the side opposite to the LHCb detector. 

%%%%%%%%%%%%%%%%%%%%%%%%%%%%%%%%%%%%%%%%%%%%%%%%%%%
\subsection{TimePix radiation monitors}\label{sc:timepix}

The only non-passive MoEDAL sub-detector is an array of TimePix pixel devices distributed throughout the MoEDAL cavern, forming a real-time radiation monitoring system of HI beam-related backgrounds. The operation in time-over-threshold mode allows a 3D mapping of the charge spreading in the volume of the silicon sensor, thus differentiating between various particles species from mixed radiation fields and measuring their energy deposition.

%%%%%%%%%%%%%%%%%%%%%%%%%%%%%%%%%%%%%%%%%%%%%%%%%%%
%%%%%%%%%%%%%%%%%%%%%%%%%%%%%%%%%%%%%%%%%%%%%%%%%%%
\section{Magnetic monopoles}\label{sc:mm}

The MoEDAL detector is designed to fully exploit the energy-loss mechanisms of magnetically charged particles~\cite{Dirac1931kp,Diracs_idea,tHooft-Polyakov}  in order to optimise its potential to discover these messengers of new physics. There are various theoretical scenarios in which magnetic charge would be produced  at the LHC~\cite{Acharya:2014nyr}: (light) 't Hooft-Polyakov monopoles~\cite{tHooft-Polyakov,Vento2013jua}, electroweak monopoles~\cite{ewk}, global monopoles~\cite{vilenkin} and monopolium~\cite{Diracs_idea,khlopov,Monopolium,Monopolium1}. Magnetic monopoles that carry a non-zero magnetic charge and dyons possessing both magnetic and electric charge are predicted by many theories including grand-unified and superstring theories~\cite{Rajantie,Kephart:2017esj}. 
 
A possible explanation for the non-observation of monopoles so far is Dirac's proposal~\cite{Dirac1931kp,Diracs_idea,khlopov} that monopoles are not seen freely because they form a bound state called \emph{monopolium}~\cite{Monopolium,Monopolium1,Epele0} being confined by strong magnetic forces. Monopolium is a neutral state, difficult to detect directly at a collider detector, although its decay into two photons would give a rather clear signal for ATLAS and CMS~\cite{Epele}, which however would not be visible in MoEDAL. Nevertheless the LHC radiation detector systems can be used to detect final-state protons $pp\to pXp$ exiting the LHC beam vacuum chamber at locations determined by their fractional momentum losses~\cite{risto}. Such technique would be appealing for detecting monopolia. 

%%%%%%%%%%%%%%%%%%%%%%%%%%%%%%%%%%%%%%%%%%%%%%%%%%%%%%%%%%%%%%%%%%%%%%%%%%%%%%%%%%%%%%%%%%%%%%%%%%%%%%
\section{Searches for monopoles in MoEDAL with MMTs}\label{sc:lightsearch}

The high magnetic charge of a monopole ---being at least one Dirac charge $g_{\rm D} = 68.5 e$--- implies a strong magnetic dipole moment, which may result in strong binding of the monopole with the nuclei of the aluminium MMTs. In such a case, the presence of a monopole trapped in an MMT bar would de detected through a non-zero \emph{persistent current}, defined as the difference between the SQUID currents before and after its passage through the sensing coil. 

\begin{figure}[ht]
  \includegraphics[width=0.46\textwidth]{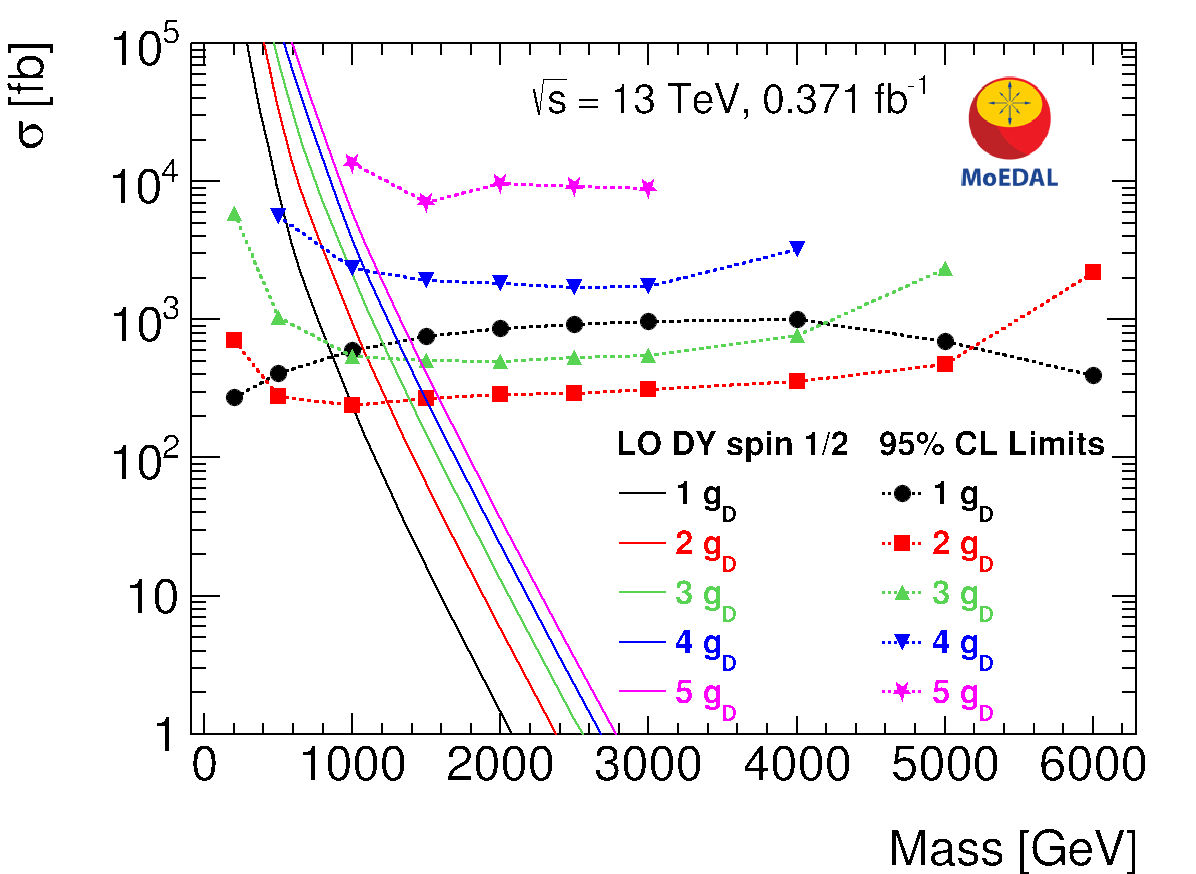}
  \hspace{0.02\textwidth}%
  \includegraphics[width=0.52\textwidth]{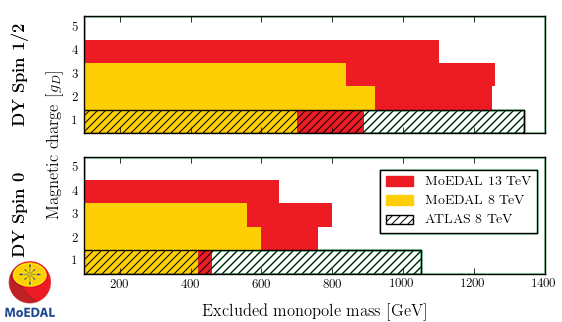}
  \caption{Left: Cross-section upper limits at 95\% C.L.\ for DY monopole production as a function of mass for spin-1/2 monopoles~\cite{MMT13TeV}. Right: Excluded monopole masses for DY production for spin-$1/2$ (top) and spin-$0$ (bottom) monopoles. The MoEDAL results obtained at 8~TeV~\cite{MMT8TeV} and 13~TeV~\cite{MMT13TeV} are superimposed on the ATLAS 8-TeV limits~\cite{atlas8tev}.}
\label{fg:limits}
\end{figure}

The Run-2 MMT configuration was analysed and no magnetic charge  $>0.5g_{\rm D}$ was detected in any of the exposed samples when passed through the ETH Zurich SQUID. Hence cross-section limits are obtained for Drell-Yan (DY) pair production of spin-1/2 and spin-0 monopoles for $1g_{\rm D}\leq|g|\leq 5g_{\rm D}$ at 13~TeV~\cite{MMT13TeV}, as shown in Fig.~\ref{fg:limits} (left), improving previous bounds set by MoEDAL at 8~TeV~\cite{MMT8TeV}. However, the large monopole-photon coupling invalidates any perturbative treatment of the cross-section calculation and hence any result based on the latter is only indicative. This situation may be resolved if thermal production in heavy-ion collisions ---that does not rely on perturbation theory--- is considered~\cite{Gould:2017zwi}.

Under the assumption of Drell-Yan cross sections, mass limits are derived for $g_{\rm D}\leq|g|\leq4g_{\rm D}$ at the LHC, complementing ATLAS results~\cite{atlas7tev,atlas8tev}, which placed limits for monopoles with magnetic charge $|g|\leq1.5 g_{\rm D}$ (c.f.\ Fig.~\ref{fg:limits}, right). The ATLAS bounds are better that the MoEDAL ones for $|g|=1 g_{\rm D}$ due to the higher luminosity delivered in ATLAS and the loss of acceptance in MoEDAL for small magnetic charges. On the other hand, higher charges are difficult to be probed in ATLAS due to the limitations of the level-1 trigger deployed for such searches. Limits on monopole production cross sections set by various colliders are presented in Ref.~\cite{Rajantie}, while general limits including searches in cosmic radiation are reviewed in Ref.~\cite{patrizii}. 

%%%%%%%%%%%%%%%%%%%%%%%%%%%%%%%%%%%%%%%%%%%%%%%%%%%
%%%%%%%%%%%%%%%%%%%%%%%%%%%%%%%%%%%%%%%%%%%%%%%%%%%
\section{Summary and outlook}\label{sc:summary}

MoEDAL extends considerably the LHC reach in the search for (meta-)stable HI particles. The latter are predicted in a variety of theoretical models and include: magnetic monopoles, SUSY long-lived spartners, quirks, strangelets, Q-balls, etc~\cite{Acharya:2014nyr,creta2016}. The MoEDAL Collaboration is preparing new analyses with more Run~2 data, with other detectors (NTDs) and with a large variety of interpretations involving not only magnetic but also electric charges.

%%%%%%%%%%%%%%%%%%%%%%%%%%%%%%%%%%%%%%%%%%%%%%%%%%%
%%%%%%%%%%%%%%%%%%%%%%%%%%%%%%%%%%%%%%%%%%%%%%%%%%%
\section*{Acknowledgments}

The author acknowledges support by the Spanish MINEICO under the project FPA2015-65652-C4-1-R, by the Generalitat Valenciana through the MoEDAL-supporting agreement CON.21.2017-09.02.03, by the CSIC CT Incorporation Program 201650I002, by the Severo Ochoa Excellence Centre Project SEV 2014-0398, and by a 2017 Leonardo Grant for Researchers and Cultural Creators, BBVA Foundation.

% ****************************************************************************
% BIBLIOGRAPHY AREA
% ****************************************************************************

\begin{footnotesize}

\end{footnotesize}

% ****************************************************************************
% END OF BIBLIOGRAPHY AREA
% ****************************************************************************

\end{document}